\documentclass[aps,pre,twocolumn,groupedaddress,showpacs,floatfix]{revtex4}
\usepackage{graphicx}
\usepackage{pst-all}
\usepackage{color}
\usepackage{amsmath}
\newcommand{\comment}[1]{}

\begin{document}
\title{Scaling in critical random Boolean networks}
\author{Viktor Kaufman, Tamara Mihaljev and Barbara Drossel}
\affiliation{Institut f\"ur Festk\"orperphysik,  TU Darmstadt,
Hochschulstra\ss e 6, 64289 Darmstadt, Germany }
\date{\today}
\begin{abstract}
We derive mostly analytically the scaling behavior of the number of
nonfrozen and relevant nodes in critical Kauffman networks (with two
inputs per node) in the thermodynamic limit. By defining and analyzing
a stochastic process that determines the frozen core we can prove that
the mean number of nonfrozen nodes scales with the network size $N$ as
$N^{2/3}$, with only $N^{1/3}$ nonfrozen nodes having two nonfrozen
inputs. We also show the probability distributions for the numbers of these
nodes. Using a different stochastic process, we determine the scaling
behavior of the number of relevant nodes. Their mean number increases for
large $N$ as $N^{1/3}$, and only a finite number of relevant nodes
have two relevant inputs. It follows that all relevant components
apart from a finite number are simple loops, and that the mean number
and length of attractors increases faster than any power law with
network size.
\end{abstract}
\maketitle

\section{Introduction}
\label{intro}

Random Boolean networks are often used as generic models for the dynamics of
complex systems of interacting entities, such as social and economic
networks, neural networks, and gene or protein interaction networks
\cite{kauffman:random}. The simplest and most widely studied of these
models was introduced in 1969 by Kauffman \cite{kauffman:metabolic} as
a model for gene regulation.  The system consists of $N$ nodes, each
of which receives input from $K$ randomly chosen other nodes. The
network is updated synchronously, the state of a node at time step $t$
being a Boolean function of the states of the $K$ input nodes at the
previous time step, $t-1$.  The Boolean updating functions are
randomly assigned to every node in the network, and together with the
connectivity pattern they define the realization of the network. For
any initial condition, the network eventually settles on a periodic
attractor. 
Of special interest are \emph{critical} networks, which lie at the boundary 
between a frozen phase and a chaotic phase \cite{derrida:random,derrida:phase}.  
In the frozen phase, a perturbation at one node propagates during one time
step on an average to less than one node, and the attractor lengths
remain finite in the limit $N\to \infty$. In the chaotic phase, the
difference between two almost identical states increases exponentially
fast, because a perturbation propagates on an average to more than one
node during one time step \cite{aldana-gonzalez:boolean}.

The nodes of a critical network can be classified according to their
dynamics on an attractor. First, there are nodes that are frozen on
the same value on every attractor. Such nodes give a constant input to
other nodes and are otherwise irrelevant. They form the \emph{frozen
core} of the network. Second, there are nodes whose outputs go only to
irrelevant nodes. Though they may fluctuate, they are also classified
as irrelevant since they act only as slaves to the nodes determining
the attractor period. Third, the \emph{ relevant nodes} are the nodes
whose state is not constant and that control at least one relevant
node. These nodes determine completely the number and period of
attractors. If only these nodes and the links between them are
considered, these nodes form loops with possibly additional links and
chains of relevant nodes within and between loops.  The recognition of
the relevant elements as the only elements influencing the asymptotic
dynamics was an important step in understanding the attractors of
Kauffman networks.  The behavior of the frozen core was first studied
by Flyvbjerg \cite{flyvbjerg:order}. Then, in an analytical study of
$K=1$ networks Flyvbjerg and Kjaer \cite{flyvbjerg:exact} introduced
the concept of relevant elements (though without using this name). The
definition of relevant elements that we are using here was given by
Bastolla and Parisi \cite{bastolla:relevant,bastolla:modular}. They
gained insight into the properties of the attractors of the critical
networks by using numerical experiments based on the modular structure
of the relevant elements. Finally, Socolar and Kauffman
\cite{socolar:scaling} found numerically that for critical $K=2$
networks the mean number of nonfrozen nodes scales as $N^{2/3}$, and
the mean number of relevant nodes scales as $N^{1/3}$. The same result
is hidden in the analytical work on attractor numbers by Samuelsson
and Troein \cite{samuelsson:superpolynomial}, as was shown in
\cite{drossel:onnumber}.

In this work, we go a step further by deriving these power laws
analytically for a more general class of networks, and by showing the
asymptotic probability distribution of nonfrozen and relevant nodes in
terms of scaling variables. We also obtain results for the number of
nonfrozen nodes with two nonfrozen inputs, and for the number of
relevant nodes with two relevant inputs.  The outline of this paper is
the following.  In the next section we define the class of networks
that we are investigating. In Section \ref{process}, we introduce a
stochastic process that determines the frozen core of the network
starting from the nodes whose outputs are entirely independent of
their inputs. Then, in Section \ref{fluctuations}, we analyze the
Langevin and Fokker-Planck equations that correspond to this
stochastic process and that lead to the scaling behavior of the number of
nonfrozen nodes. In order to identify the relevant nodes among the
nonfrozen ones, we introduce in Section \ref{relevant} another
stochastic process. This process also enables us to find their scaling
behavior. Finally, we discuss in the last section the implications of
our results.

\section{Critical $K=2$ networks}
\label{definition}

The networks we are studying in this paper are the $K=2$ critical
networks. In these networks each node has 2 randomly chosen
inputs. The 16 possible update functions are shown in table
\ref{tab1}.

\begin{table} \begin{center}
\begin{tabular}{|c||c|c||c|c|c|c||c|c|c|c|c|c|c|c||c|c|}\hline In&
\multicolumn{2}{|c||}{$\mathcal{F}$}&
\multicolumn{4}{|c||}{${\mathcal{C}}_1$}&
\multicolumn{8}{|c||}{${\mathcal{C}}_2$}&
\multicolumn{2}{|c|}{$\mathcal{R}$}\\\hline
00&1&0&0&1&0&1&1&0&0&0&0&1&1&1&1&0\\
01&1&0&0&1&1&0&0&1&0&0&1&0&1&1&0&1\\
10&1&0&1&0&0&1&0&0&1&0&1&1&0&1&0&1\\
11&1&0&1&0&1&0&0&0&0&1&1&1&1&0&1&0\\\hline \end{tabular} \end{center}
\caption{The 16 update functions for nodes with 2 inputs. The first
column lists the 4 possible states of the two inputs, the other
columns represent one update function each, falling into four
classes.}  \label{tab1} \end{table} The update functions fall into
four classes \cite{aldana-gonzalez:boolean}. In the first class,
denoted by $\mathcal{F}$, are the frozen functions, where the output
is fixed irrespectively of the input. The class ${\mathcal{C}}_1$
contains those functions that depend only on one of the two inputs,
but not on the other one. The class ${\mathcal{C}}_2$ contains the
remaining canalizing functions, where one state of each input fixes
the output. The class $\mathcal{R}$ contains the two reversible update
functions, where the output is changed whenever one of the inputs is
changed. Critical networks are those where a change in one node
propagates to one other node on an average. A change propagates with
probability $1/2$ to a node that has a canalizing update function
$\mathcal{C}_1$ or $\mathcal{C}_2$, with probability zero to a node
that has a frozen update function, and with probability 1 to a node
that has a reversible update function. Consequently, if the frozen and
reversible update functions are chosen with equal probability, the
network is critical. Usually, only those models are considered where
all 16 update functions receive equal weight. We here consider the
larger set of models where the frozen and reversible update functions
are chosen with equal (and nonzero) probability, and where the
remaining probability is divided between the $\mathcal{C}_1$ and
$\mathcal{C}_2$ functions. Those networks that contain only
$\mathcal{C}_1$ functions are different from the remaining ones.
Since all nodes respond only to one input, the link to the second
input can be cut, and we are left with a critical $K=1$ network, which
was already discussed in
\cite{flyvbjerg:exact,drossel:number,drossel:onnumber} and will not be
discussed here. All the other models, where the weight of the
$\mathcal{C}_1$ functions is smaller than 1, fall into the same class
\cite{drossel:onnumber}. The treatment presented in the following, is
based on the existence of nodes with frozen functions, and it
therefore applies to all critical models with a nonzero fraction of
frozen functions. Networks with only canalyzing functions have to be
discussed separately. 

Let $N_f$ be the number of nodes with a frozen function, $N_r$ the
number of nodes with a reversible function and $N_{c_1}$ and $N_{c_2}$
the number of nodes with a $\mathcal{C}_1$ and a $\mathcal{C}_2$
function. We define the systems we are going to consider through
parameters $\alpha = N_{c_1}/N$, $\beta = N_r/N = N_f/N$, $\gamma =
N_{c_2}/N$. These parameters give the fraction of each type of nodes
in the network. In the next two sections, we determine the properties
of the frozen core in the large $N$ limit by starting from the nodes
with a frozen function.

\section{A stochastic process that leads to the frozen core}
\label{process}

We consider the ensemble of all networks of size $N$ and with fixed
parameters $\alpha, \beta, \gamma$. All nodes with a frozen update
function are certainly part of the frozen core. We now construct the
frozen core by determining stepwise all those nodes that become frozen
due to the influence of a frozen node. In the language of
\cite{socolar:scaling}, this process determines the ``clamped''
nodes. Initially, we place the nodes in four containers labelled
$\mathcal{F}$, $\mathcal{C}_1$, $\mathcal{C}_2$, and
$\mathcal{R}$. These containers contain $N_f$, $N_{c_1}$, $N_{c_2}$,
and $N_r$ nodes initially. Since these numbers change during our
stochastic process, we denote the initial values as $N_f^{ini}$,
$N_{c_1}^{ini}$, $N_{c_2}^{ini}$, and $N_r^{ini}$, and the total
number of nodes as $N^{ini}$.  We treat the nodes in container
$\mathcal{C}_1$ as nodes with only one input and with the update
functions ``copy'' or ``invert''.  The contents of the containers will
change with time. The ``time'' we are defining here is not the real
time for the dynamics of the system. Instead, it is the time scale for
a stochastic process that we use to determine the frozen core. During
one time step, we remove one node from the container $\mathcal{F}$ and
determine all those nodes, to which this node is an input. A node in
container $\mathcal{C}_1$ chooses this node as an input with
probability $1/N$. It then becomes a frozen node. We therefore move
each node of container $\mathcal{C}_1$ with probability $1/N$ into the
container $\mathcal{F}$. A node in container $\mathcal{C}_2$ chooses
the selected frozen node as an input with probability $2/N$. With
probability $1/2$, it then becomes frozen, because the frozen node is
with probability $1/2$ in the state that fixes the output of a
$\mathcal{C}_2$-node. If the $\mathcal{C}_2$-node does not become
frozen, it becomes a $\mathcal{C}_1$-node. We therefore move each node
of container $\mathcal{C}_2$ during the first time step with
probability $1/N$ into the container $\mathcal{F}$, and with
probability $1/N$ into the container $\mathcal{C}_1$. Finally, a node
in container $\mathcal{R}$ chooses the selected frozen node as an
input with probability $2/N$ and becomes a $\mathcal{C}_1$-node. We
therefore move each node of container $\mathcal{R}$ during the first
time step with probability $2/N$ into the container $\mathcal{C}_1$.
In summary, the total number of nodes, $N$, decreases by one during
one time step, since we remove one node from container $\mathcal{F}$,
and some nodes move to a different container. The removed nodes are
those frozen nodes for which we already have determined whose input
they are. Then, we take the next frozen node out of container
$\mathcal{F}$ and determine its effect on the other nodes. We repeat
this procedure until we cannot continue because either container
$\mathcal{F}$ is empty, or because all the other containers are
empty. If container $\mathcal{F}$ becomes empty, we are left with the
nonfrozen nodes. We shall see below that most of the remaining nodes
are in container $\mathcal{C}_1$, with the proportion of nodes left in
containers $\mathcal{C}_2$ and $\mathcal{R}$ vanishing in the limit
$N^{ini}\to \infty$. Then, the nonfrozen nodes can be connected to a
network by choosing the input(s) to every node at random from the
other remaining nodes. If all containers apart from container
$\mathcal{F}$ are empty at the end, the entire network becomes
frozen. This means that the dynamics of the network go to the same
fixed point for all initial conditions.

Let us first describe this process by deterministic equations that
neglect fluctuations around the average change of the number of nodes
in the different containers. As long as all containers contain large
numbers of nodes, these fluctuations are negligible, and the
deterministic description is appropriate. The average change of the
node numbers in the containers during one time step is
\begin{eqnarray}
\Delta N_r &=& - \frac{2N_r}{N}\nonumber \\
\Delta N_{c_2} &=& - \frac{2N_{c_2}}{N}\nonumber \\
\Delta N_{c_1} &=& - \frac{N_{c_1}}{N} + \frac{N_{c_2}}{N} +
\frac{2N_r}{N} \label{Delta}\\
\Delta N_f &=& -1 +  \frac{N_{c_1}}{N} + \frac{N_{c_2}}{N}\nonumber \\
\Delta N &=& -1\nonumber
\end{eqnarray}
The number of nodes in the containers, $N$, can be used instead of the
time variable, since it decreases by one during each step. The
equation for $N_r$ can then be solved by going from a difference
equation to a differential equation,
$$\frac {\Delta N_r}{\Delta N} \simeq \frac {d N_r}{d N} =  -
\frac{2N_r}{N}\, ,$$
which has the solution
\begin{equation}
N_r = N^2 \frac{N_r^{ini}}{(N^{ini})^2}\, . \label{Nr}
\end{equation}
Similarly, we find 
\begin{eqnarray}
N_{c_2} &=& N^2 \frac{N_{c_2}^{ini}}{(N^{ini})^2}\nonumber \\
N_f &=& N\frac{N_f^{ini}-N_r^{ini}}{N^{ini}}+N^2 \frac{N_r^{ini}}{(N^{ini})^2} \nonumber \\
N_{c_1} &=&  N \frac{N_{c_1}^{ini}+N_{c_2}^{ini}+2N_r^{ini}}{N^{ini}}- 2 N^2 \frac{N_r^{ini}+N_{c_2}^{ini} }{(N^{ini})^2}
\, . \label{det}
\end{eqnarray}
For $N_f^{ini}<N_r^{ini}$, we obtain $N_f=0$ at a nonzero value of
$N$, and the number of nonfrozen nodes is proportional to
$N^{ini}$. We are in the chaotic phase. For $N_f^{ini}>N_r^{ini}$, the
values $N_r$ and $N_{c_2}$ will sink below 1 when $N$ becomes of the
order $\sqrt{N^{ini}}$. For smaller $N$, there are only $\mathcal{F}$
and $\mathcal{C}_1$ nodes left, and the second term contributing to
$N_f$ and $N_{c_1}$ in (\ref{det}) can be neglected compared to the
first one. When $N_f$ falls below 1, there remain
$N_{c_1}=\frac{N_{c_1}^{ini}+N_{c_2}^{ini}+2N_r^{ini}}{N_f^{ini}-N_r^{ini}}$
nodes of type $\mathcal{C}_1$. The network is essentially frozen, with
only a finite number of nonfrozen nodes in the limit $N^{ini} \to
\infty$. If we now choose the inputs for these nodes, we obtain simple
loops with trees rooted in the loops. This property of the frozen
phase was also found in \cite{socolar:scaling}.

For the critical networks that this paper focuses on, we have 
$N_f^{ini}=N_r^{ini}= \beta N^{ini}$, and the stochastic process stops
at $N_f = 1 = \beta {N^2}/{N^{ini}}$. This means that 
\begin{equation}
N^{end} = \sqrt{\frac{N^{ini}}{\beta}}\, .
\end{equation}
The number of nonfrozen nodes would scale with the square root of the
network size if the deterministic approximation to the stochastic
process was exact. We shall see below that including fluctuations
changes the exponent from $1/2$ to $2/3$.  The final number of
$\mathcal{C}_2$-nodes for the deterministic process for the critical
networks is $\gamma/\beta$, which is independent of network size, and
the final number of $\mathcal{R}$-nodes vanishes due to $N_r =
N_f$. We shall see below that the fluctuations change these two
results to a $(N^{ini})^{1/3}$-dependence.

Introducing $n = N/N^{ini}$ and $n_j = N_j/N^{ini}$ for $j=r,f,c_1,c_2$, 
equations (\ref{det}) simplify to (using $N_r^{ini} = N_f^{ini}$)
\begin{eqnarray}
n_r &=& \beta n^2 = n_f\nonumber\\
n_{c_2} &=& \gamma n^2 \nonumber\\
n_{c_1} &=& n-2\beta n^2 - \gamma n^2 
\, .\nonumber
\end{eqnarray}
This means that our stochastic process remains invariant (in the
deterministic approximation) when the initial number of nodes in the
containers and the time unit are all multiplied by the same factor.
For small $n$, the majority of nodes are in container $\mathcal{C}_1$,
since $n_{c_1}= n - \mathcal{O}(n^2)$. Now, if we choose a
sufficiently large $N^{ini}$, $n$ reaches any given small value while
$N_f=N_r=\beta n^2N^{ini} $ is still large enough for a deterministic
description.  We can therefore assume that for sufficiently large
networks $N_f/N=\beta n$ becomes small before the effect of the noise
becomes important. This assumption will simplify our calculations
below.

\section{The effect of fluctuations}
\label{fluctuations}

The number of nodes in container $\mathcal{C}_1$ that choose a given
frozen node as an input is Poisson distributed with a mean $N_{c_1}/N$
and a variance $N_{c_1}/N$. We now assume that $n$ is small at the
moment where noise becomes important, i.e., that the variance of the
noise $N_{c_1}/N = n_{c_1}/n = 1-(2\beta+\gamma)n = 1-\mathcal{O}(n)$
is unity. The number of nodes in containers $\mathcal{C}_2$ and
$\mathcal{R}$ that choose a given frozen node as an input is Poisson
distributed with a mean and a variance $2(N_{c_2} + N_r)/N$. The
fluctuation around the mean can be neglected as this noise term is
very small compared to $N_r $ and $N_{c_2}$, the final values of
which are large for sufficiently large $N^{ini}$.  We therefore obtain
the stochastic version of equations (\ref{Delta})
\begin{eqnarray}
\Delta N_r &=& - \frac{2N_r}{N}\nonumber \\
\Delta N_{c_2} &=& - \frac{2N_{c_2}}{N}\nonumber \\
\Delta N_f &=&-\frac{N_{r}}{N}-\frac{N_{f}}{N}+\xi \nonumber \\
\Delta N &=& -1\label{delta2}
\end{eqnarray}
The random variable $\xi$ has zero mean and unit variance. As long as
the $n_j$ change little during one time step, we can summarize a large
number $T$ of time steps into one effective time step, with the noise
becoming Gaussian distributed with zero mean and variance $T$. Exactly
the same process would result if we summarized $T$ time steps of a
process with Gaussian noise of unit variance. For this reason, we can
choose the random variable $\xi$ to be Gaussian distributed with unit
variance.

Compared to the deterministic case, the equations for $N_r$ and $
N_{c_2}$ are unchanged, and we have again 
$N_r = N^2 N_r^{ini}/(N^{ini})^2$ and 
$N_{c_2} = N^2 N_{c_2}^{ini}/(N^{ini})^2$. Inserting the
solution for $N_r$ into the equation for $N_f$, we obtain
\begin{equation}
\frac{dN_f}{dN} = \frac{N_{f}}{N} +  \frac{\beta N}{N^{ini}}+\xi\label{langevin}
\end{equation}
with the step size $dN=1$ and $\langle \xi^2\rangle = 1$. (In the
continuum limit $dN\to 0$ the noise correlation becomes $\langle
\xi(N)\xi(N')\rangle = \delta(N-N')$).  This is a Langevin-equation,
and we will now derive the corresponding Fokker-Planck-equation. Let
$P(N_f,N)$ be the probability that there are $N_f$ nodes in container
$\mathcal{R}$ at the moment where there are $N$ nodes in total in the
containers. This probability depends on the initial node number
$N_{ini}$, and on the parameter $\beta$. The sum
$$\sum_{N_f=1}^\infty P(N_f,N) \simeq\int_0^\infty P(N_f,N) dN_f $$ is the
probability that the stochastic process is not yet finished, i.e. the
probability that $N_f$ has not yet reached the value 0 at the moment
where the total number of nodes in the containers has decreased to the
value $N$. Since systems that have reached $N_f=0$ are removed from
the ensemble, we have to impose the absorbing boundary condition
$P(0,N)=0$.  Let $g(\Delta N_f|N_f,N)$ denote the probability that
$N_f$ decreases by $\Delta N_f$ during the next step, given the values
of $N_f$ and $N$.

We have
\begin{eqnarray}
&&P(N_f,N-1) = \nonumber\\&&\int_0^\infty P(N_f + \Delta N_f, N)g(\Delta N_f|N_f+\Delta N_f,N) d(\Delta
N_f)\nonumber\\
&&=  \int_0^\infty\Bigl[ P(N_f, N)g(\Delta N_f|N_f,N) \nonumber\\&&\quad + \frac{\partial}{\partial
    N_f}(P(N_f, N)g(\Delta N_f|N_f,N))\Delta N_f   \nonumber\\
 &&\quad+\frac{\partial^2}{2\partial^2   N_f^2}(P(N_f, N)g(\Delta N_f|N_f,N))(\Delta N_f)^2 \nonumber\\&&\quad+ \ldots\Bigr] d(\Delta
N_f)\nonumber\\
&&= P(N_f, N) + \frac{\partial}{\partial
    N_f} (P(N_f, N) \langle \Delta N_f \rangle) + \nonumber\\&&\quad \frac{\partial^2}{2\partial
   N_f^2}(P(N_f, N) \langle (\Delta N_f)^2 \rangle) +
\ldots\nonumber
\end{eqnarray}
The mean change  $\langle \Delta N_f \rangle$ during one step is
$\langle \Delta N_f \rangle = \frac{N_{f}}{N} +  \frac{\beta
  N}{N^{ini}} $, and the mean square change is 
$\langle (\Delta N_f)^2 \rangle \simeq 1$. 

This gives the Fokker-Planck equation for our stochastic process
\begin{equation}
-\frac{\partial P}{\partial N} = \frac{\partial}{\partial
    N_f} \left( \frac{N_{f}}{N} +  \frac{\beta
  N}{N^{ini}} \right)P + \frac 1 2 \frac{\partial^2 P}{\partial
   N_f^2}\, . \label{FP}
\end{equation}
We introduce the variables 
\begin{equation}
x = \frac {N_f}{\sqrt{N}} \hbox{ and } y = \frac{N}{(N^{ini}/\beta)^{2/3}}\label{defxy}
\end{equation}
and the function $f(x,y)= (N^{ini}/\beta)^{1/3}P(N_f,N)$. We will see
below that $f(x,y)$ does not depend explicitely on the parameters
$N^{ini}$ and $\beta$ with this definition. The Fokker-Planck equation
then becomes
\begin{equation}
y\frac{\partial f}{\partial y} +f+\left(\frac x 2 + y^{3/2}\right)
    \frac{\partial f}{\partial
    x} + \frac 1 2 \frac{\partial^2 f}{\partial
   x^2} = 0\, . \label{FP2}
\end{equation}
Let $W(N)$ denote the probability that $N$ nodes are left at the
moment where $N_f$ reaches the value zero. It is
\begin{eqnarray*}
W(N)  &=& \quad \int_0^\infty P(N_f,N)dN_f -  \int_0^\infty
P(N_f,N-1)dN_f \, .
\end{eqnarray*}
Consequently,
\begin{eqnarray}
W(N) &=& \frac{\partial}{\partial N}\int_0^\infty P(N_f,N)
dN_f\nonumber\\
&=& (N^{ini}/\beta)^{-1/3}\frac{\partial}{\partial N}\sqrt N \int_0^\infty f(x,y)dx
\nonumber\\
&=&  (N^{ini}/\beta)^{-2/3} \frac{\partial}{\partial y}\sqrt y \int_0^\infty f(x,y)dx
\nonumber\\
&\equiv&  (N^{ini}/\beta)^{-2/3}  G(y)\label{w}
\end{eqnarray}
with a scaling function $G(y)$. $W(N)$ must be a normalized function,
$\int_0^\infty W(N)dN = \int_0^\infty G(y) dy = 1$.  This condition is
independent of the parameters of the model, and therefore $G(y)$ and
$f(x,y)$ are independent of them, too, which justifies our choice of
the prefactor in the definition of $f(x,y)$. By integrating equation
(\ref{FP2}) over $x$ from 0 to infinity and by using
$f(0,y)=f(\infty,y)=0$ we obtain 
$$\sqrt y \frac{\partial}{\partial y} \sqrt y \int_0^\infty f dx - \frac 1 2
\left. \frac{\partial f}{\partial x}\right|_{x=0} =0\, ,$$
which gives us a second relation between $f(x,y)$
and $G(y)$:
\begin{equation}
\sqrt y G(y) = \frac 1 2 \left. \frac{\partial f}{\partial x}\right|_{x=0}\, .
\label{secondg}
\end{equation}

\begin{figure}
\includegraphics[width=0.4\textwidth]{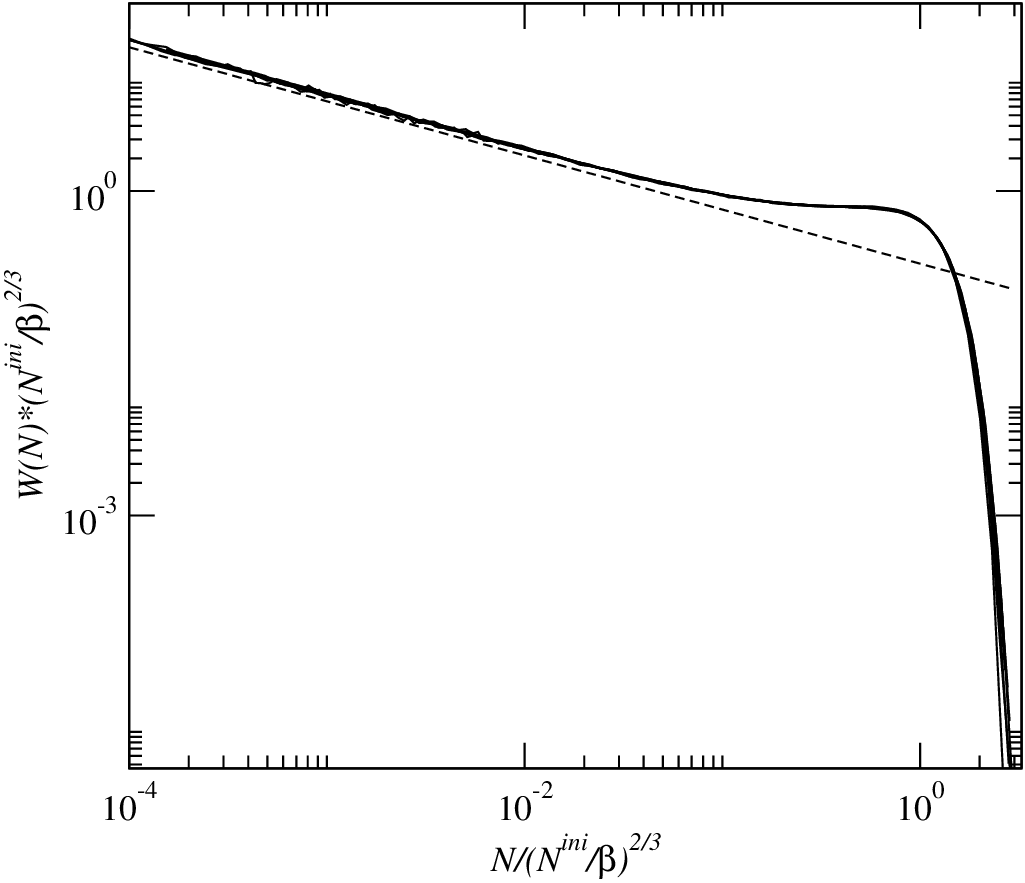}
\caption{
The function 
$W(N)(N^{ini}/\beta)^{2/3}$ 
vs $N/(N^{ini}/\beta)^{2/3}$ 
for $\beta = 0.25$ and 
$N^{ini}=2^{16},2^{17},2^{18},2^{19},2^{20},2^{21}$. 
Furthermore, the graph contains a curve with $\beta=0.125, N=2^{16}$ 
and a curve with $\beta=0.5, N=2^{16}$. The curves all collapse,
confirming the existence of a 
scaling function $G(y)$. The dashed line is a power law $\sim 1/\sqrt{N}$.
}
\label{fig1}
\end{figure}

The mean number of nonfrozen nodes is
\begin{equation}\bar N = \int_0^\infty N W(N)dN =  (N^{ini}/\beta)^{2/3}  \int_0^\infty G(y)
ydy\, ,\label{barN}
\end{equation}
which is proportional to $ (N^{ini}/\beta)^{2/3}$. 
We did not succeed in extracting an explicit expression for the
function $G(y)$. It can be determined by running the stochastic
process described by the  equations (\ref{delta2}) on the computer. 
The result is shown in Figure \ref{fig1}, and an almost perfect fit to
this result is given by 
\begin{equation} G(y) \simeq 0.25 e^{-y^3/2}(1-0.5\sqrt{y} + 3y)/\sqrt{y}\, .
\label{gfit}
\end{equation}

\begin{figure}
\includegraphics[width=0.4\textwidth]{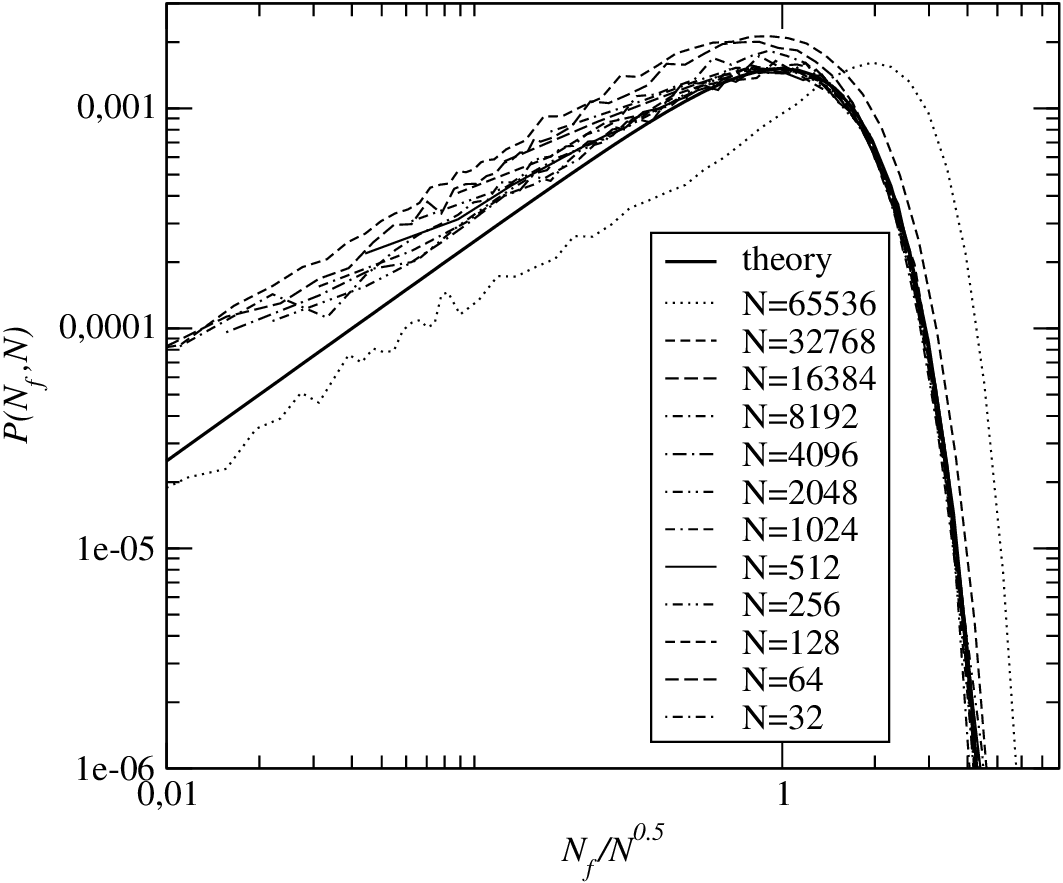}
\caption{$P(N_f,N)$ vs $N_f/\sqrt{N}$ for $N^{ini}= 2^{21}$ and
$\beta=1/4$ for different $N$. The thick solid line is the theoretical
result Eq.~(\ref{fsmally}), which is  approached in the limit of
 small $N/(N^{ini})^{2/3}$. }
\label{fig2}
\end{figure}

For small $y$, the data show a power law $G(y) \propto y^{-1/2}$. We
can obtain this power law analytically by solving the Fokker-Planck
equation (\ref{FP2}) in the limit of small $y$. In this limit, the
term proportional to $y^{3/2}$ can be dropped, and we have the simpler
equation
\begin{equation}
y\frac{\partial f}{\partial y} +f+\frac x 2 
    \frac{\partial f}{\partial
    x} + \frac 1 2 \frac{\partial^2 f}{\partial
   x^2} = 0\, . \label{FP3}
\end{equation}
The general solution has the form $f(x,y)=\sum_\nu c_\nu y^\nu f_\nu(x)$,
with the functions $f_\nu$ satisfying
\begin{equation}
2(\nu+1)f_\nu + x f_\nu' + f_\nu'' = 0\, .\label{fnu}
\end{equation}
The solution is 
$$e^{\frac{x^2}{2}}f_\nu(x) =C_1H_{1+2\nu}\left(\frac x{\sqrt
2}\right)+C_2\, _1F_1\left(-\nu-\frac 1 2 ;\frac 1
2;\frac{x^2}2\right)$$ with two constants $C_1$ and $C_2$, and with
$H$ denoting the Hermitian functions, and $_1F_1$ the appropriate
hypergeometric functions. We expect $f$ to be analytical in $y$ for
small $y$, which means that $\nu = 0,1,2,\dots$. For sufficiently
small $y$, only the term $\nu=0$ contributes, and due to the absorbing
boundary condition we have $C_2=0$. We obtain therefore for small $y$
\begin{equation}
f(x,y) = c_0 xe^{-x^2/2}\, . \label{fsmally}
\end{equation}
From our numerical result (\ref{gfit}), together with (\ref{secondg}), we find  $c_0=0.5$.
Inserting  Eq.~(\ref{fsmally}) into Eq.~(\ref{w}), we obtain for small $N$
\begin{equation}
W(N)= \left(\frac{N^{ini}}{\beta}\right)^{-1/3} \frac{
  c_0}{2\sqrt{N}}\, . \label{smally}
\end{equation}
In Eq.~(\ref{fsmally}), the function $f(x,y)$ is independent of
$y$. This means that for sufficiently small $N$ the function
$P(N_f,N)$ depends only on the ratio $N_f/\sqrt N$. This is also
confirmed by our computer simulations (see Fig.~\ref{fig2}). 

We can obtain a set of solutions of Eq.~(\ref{FP2}) with the Ansatz
$f(x,y)=\sum_\nu y^\nu \tilde f_\nu(z)$ with $z=x-y^{3/2}$. The
resulting equation for $\tilde f_\nu$, is identical to Eq.~(\ref{fnu})
for $f_\nu$, which was valid for small $y$. However, an analytical
expression for the function $G(y)$ can only be given if an expansion
of the initial condition $P(N_f,N) = \delta(N_f-\beta N^{ini})$ in terms of
known solutions can be found.

The probability $W_r(N_r)$ that $N_r$ nodes are left in container 
 $\mathcal{R}$ at the moment where container $\mathcal{F}$ becomes
 empty, is obtained from the relation $$N_r = N^2
 N_r^{ini}/(N^{ini})^2\, .$$ Defining 
$$s=\frac{N_r}{(N^{ini}/\beta)^{1/3}}=y^2$$
and 
\begin{equation}
F(s) = \frac{ G(\sqrt s)}{2\sqrt s}\, ,\label{FG}
\end{equation}
 and remembering
$W(N)dN = W_r(N_r)dN_r$, we find
\begin{equation}
W_r(N_r) =   (N^{ini}/\beta)^{-1/3} F(s)\, .
\end{equation}
The mean number of nodes left in  in container 
 $\mathcal{R}$ is
\begin{eqnarray}\bar N_r &=& \int_0^\infty W_r(N_r) N_r dN_r =
 (N^{ini}/\beta)^{1/3}\int_0^\infty sF(s)ds\nonumber\\
& =&
 (N^{ini}/\beta)^{1/3} \int_0^\infty y^2G(y)dy  \,
  . 
\end{eqnarray}

\begin{figure}
\includegraphics[width=0.4\textwidth]{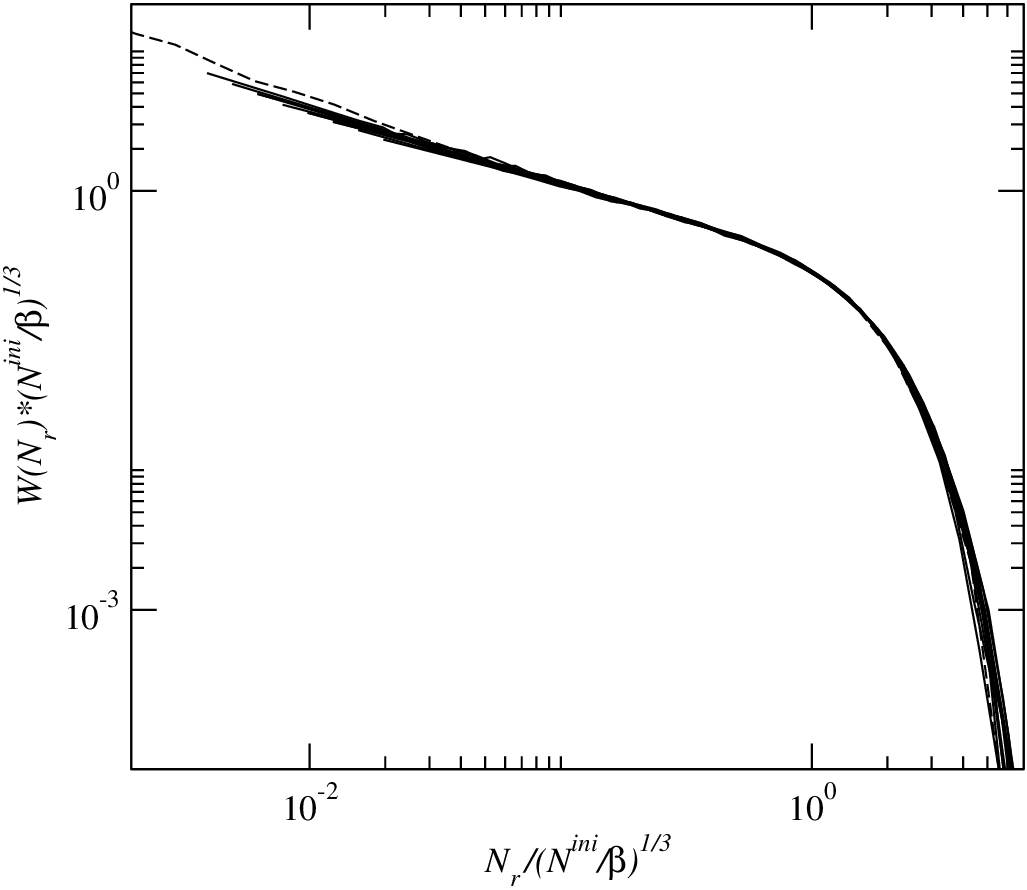}
\caption{
The function 
$W(N_r)(N^{ini}/\beta)^{1/3}$ 
vs $N_r/(N^{ini}/\beta)^{2/3}$ 
for $\beta = 0.5$ and $\beta = 0.125$ and for
$N^{ini}=2^{16},2^{17},2^{18},2^{19},2^{20},2^{21}$. 
The 12 curves converge with increasing $N$ towards an asymptotic
curve, confirming the existence of an asymptotic
scaling function $F(s)$. The dashed line shows the function $F(s)$
obtained using the data for $G(y)$ obtained from the same simulation
and Eq.~(\ref{FG}).
}
\label{fig3}
\end{figure}
The number of nodes left in container $\mathcal{C}_2$ is
$N_{c_2}=(\gamma/\beta)  N_r$. 

We thus have shown that the number of nonfrozen nodes scales with
network size $N^{ini}$ as $(N^{ini})^{2/3}$, with most of these nodes
receiving only one input from other nonfrozen nodes. The number of
nonfrozen nodes receiving two inputs from nonfrozen nodes scales 
 as $(N^{ini})^{1/3}$. We have found scaling functions that describe
 the probability distribution for these two types of nodes in the
 limit of large network size. Our next task will be to connect these
 nonfrozen nodes to a network. This is a reduced network, where all
 frozen nodes have been cut off. 

\section{Relevant nodes}
\label{relevant}

Let us start from the result obtained from the stochastic process of
the previous two sections. Each time we run this process we obtain $N$
nonfrozen nodes. Out of these, $N_r$ ($N_{c_2}$) nodes receive input
from two other nonfrozen nodes and have a reversible (canalizing
$\mathcal{C}_2$) update function. We define the parameter
\begin{equation}
a = \frac{N_r +N_{c_2} }{\sqrt N} = (1+\gamma/\beta)y^{3/2}\, ,\label{defa}
\end{equation}
which has a probability distribution $f(a)$ that is determined from
the condition $f(a)da = G(y)dy$, 
\begin{equation}
f(a) = \frac {2}{3a^{1/3}(1+\gamma/\beta)^{2/3}}G\left(\left(\frac a
{1+\gamma/\beta}\right)^{2/3}\right)\, . \label{fa}\end{equation} Just
as $G(y)$, the function $f(a)$ is the exact probability distribution
only in the thermodynamic limit $N^{ini} \to \infty$.  We determine
the relevant nodes by a stochastic process that removes iteratively
nodes that are not relevant. Each of the $N$ nonfrozen nodes chooses
its input(s) at random from the nonfrozen nodes. There are altogether
$N(1+a/\sqrt{N})$ inputs to be chosen, and consequently the nonfrozen
nodes have together $N(1+a/\sqrt{N})$ outputs. The number of outputs
of a node is Poisson distributed with the mean value
$(1+a/\sqrt{N})$. The fraction $\exp(-1-a/\sqrt{N})$ of nodes have no
output. They are the leaves of the trees of the network of nonfrozen
nodes, and we therefore know that they are not relevant. We put them
in container number 1. Their number will change during the stochastic
process that determines the relevant nodes. The other nodes are placed
in container number 2. Their number is $N_{l}$ (``labelled''), and it
will be reduced until only the relevant nodes are left. The total
number of outputs of the nodes in container 2 is initially
$N(1+a/\sqrt{N})$, while their total number of inputs is
$N(1+a/\sqrt{N})(1-\exp(-1-a/\sqrt{N}))$.  Now, we remove one node
from container 1 and connect its input(s) at random to the outputs of
the nodes in container 2. The chosen output(s) are cut off. If a node
whose output is cut off has no other output left, we move the node
from container 2 to container 1. It cannot be a relevant node since
relevant nodes influence other relevant nodes. We iterate this
procedure, until there is no node left in container 1. The nodes
remaining in container 2 are the relevant nodes. During the entire
process, the number of outputs in container 2 is identical to the
number of inputs in container 1 and 2. As long as container 1 is not
empty, there are more outputs in container 2 than inputs, and only
when the process is finished do the two numbers become identical. We
can therefore simplify the stochastic process by removing container 1
altogether. We simply have to continue cutting of outputs from nodes
in container 2 and removing nodes with no outputs, until the total
number of outputs of the nodes in container 2 has become identical to
their total number of inputs. The remaining nodes are relevant, and we
have then $N_{l}^{final} \equiv N_{rel}$. These nodes can then be
connected to a network by connecting the inputs and outputs pairwise.

In order to derive analytical results, it is useful to run this
process backwards. Starting with $N$ nodes with no outputs, adding
outputs at random will eventually generate the Poisson distribution of
the number of outputs per node that we have started with.  The reverse
stochastic process is therefore defined by the following rule: Begin
with an empty container (former container 2) and $N$ nodes outside the
container. Most of these nodes have one input, and the fraction
$a/\sqrt{N}$ have two inputs. Add an output to a randomly chosen
node. Put this node in the container.  Add another output to a randomly
chosen node (choosing every node with equal probability, whether the
node is inside or outside the container). If a node from outside the
container is chosen, put it in the container. Eventually, the total
number of outputs in the container will become larger than the total
number of inputs in the container. The container contains the relevant
nodes at the moment when the inputs equal the outputs for the last
time.

In order to show that the number of relevant nodes scales with
$\sqrt{N}$, we define a scaling variable 
$$t = \frac{N_{l}}{\sqrt{N}}\, .$$ During one step, an output is
added to nodes that are already in the container with probability
${N_{l}}/N$. Let $N_o$ count the number of outputs that have been
added to nodes in the container, i.e., $N_o = $(total number of
outputs in the container) $ - N_{l}$. Then the average rate of increase of
$N_o$ is given for sufficiently large $N$ by $$\langle\frac{dN_o}{dN_{l}}\rangle =
\frac{N_{l}}{N}\, ,$$ or
$$\langle\frac{dN_o}{dt}\rangle = t\, .$$

Let $N_i$ count the number of nodes in the container with two inputs.
Their rate of increase is $$\langle\frac{dN_i}{dN_{l}}\rangle = \frac a {\sqrt N}\, ,$$
or
$$\langle\frac{dN_i}{dt}\rangle = a\, .$$
Consequently, the probability distribution for $N_o$ is given by
\begin{equation}
P_o(N_o|t) = \frac 1 {N_o!}e^{-t^2/2}\left(\frac {t^2} 2\right)^{N_o}\, ,\label{po}
\end{equation}
and  the probability distribution for $N_i$ is given by
\begin{equation}
P_i(N_i|t) = \frac 1 {N_i!}e^{-at}\left(at\right)^{N_i}\, .\label{pi}
\end{equation}

\begin{figure}
\includegraphics*[width=0.4\textwidth]{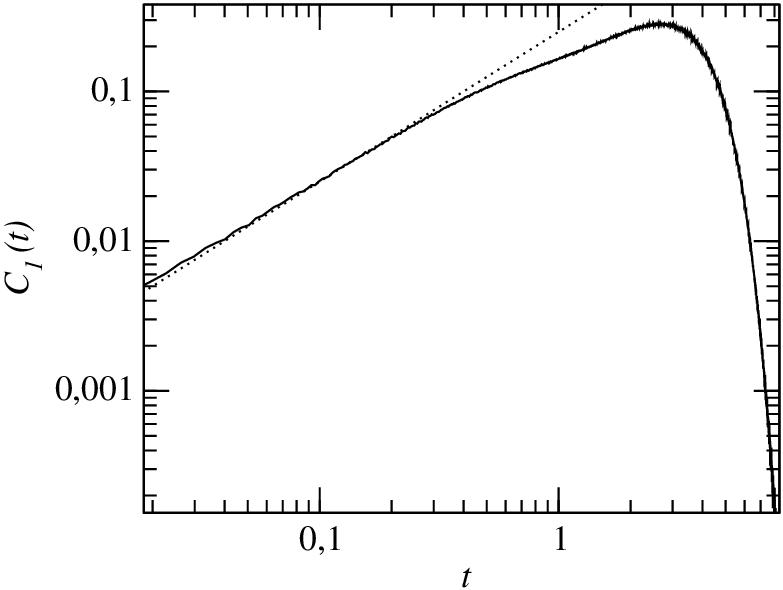} \caption{ The function
$\mathcal{C}_1(t)$ as obtained by running the stochastic process
described in this section. The dotted line corresponds to the
function $0.25 t$, which is a good fit to
$\mathcal{C}_1(t)$ for small $t$.  } \label{fig4}
\end{figure}

The stochastic process can be viewed as a random walk that steps to
the right with a rate $t$ and to the left with a rate $a$. It is
finished when $N_i=N_o$ for the last time, i.e. when the walk leaves
the origin for the last time.  We determined the probability
distribution $\mathcal{C}_a(t)$ for this last exit time
from the origin by a computer simulation. It is shown in
Fig.~\ref{fig4} for $a=1$.  For small $t$, it increases linearly in
$t$, because the probability of making a step to the right is
proportional to $t$ for small times.
For $a=0$, we can obtain an analytical result from the relation
\begin{equation}
\mathcal{C}_0(t) = -\frac{\partial P_o(0,t)}{\partial t} = te^{-t^2/2}\, .
\end{equation}

\begin{figure}
\includegraphics*[width=0.4\textwidth]{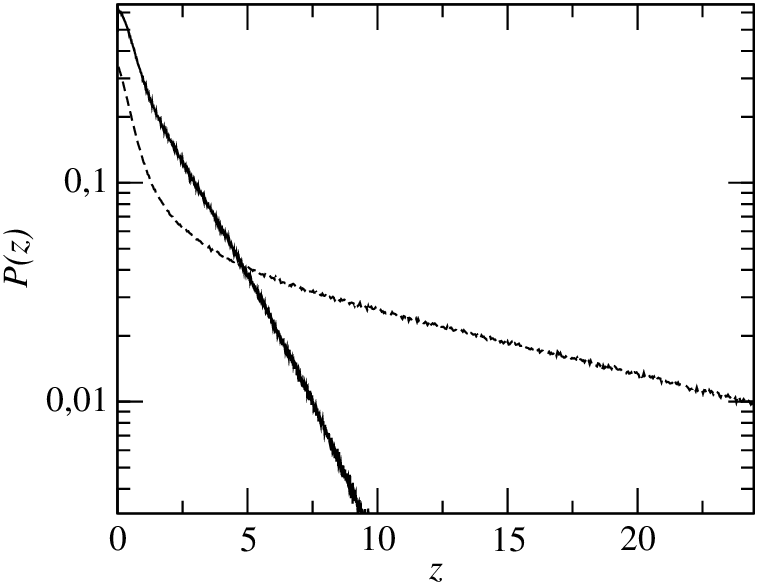} 
\caption{The function
$P(z)$ for $\gamma/\beta=0$ (solid line) and $\gamma/\beta=4$ (dashed
line). The results were obtained by running the two coupled stochastic
processes for $10^7$ samples.} 
\label{fig5}
\end{figure}

Since we were able to write the stochastic process in terms of
$t$ and $a$ alone, the probability distribution for the number of
relevant nodes depends only on the combination $N_{rel}/{\sqrt N}$ and
on the parameter $a$, 
\begin{equation}p_a(N_{rel}) dN_{rel} = \mathcal{C}_a\left(N_{rel}/{\sqrt
N}\right)dN_{rel}/{\sqrt N}\, .\label{pa}\end{equation}
The relation between $N$ and $a$ is obtained using Eq.~(\ref{defxy}) and (\ref{defa}):
$$\sqrt{N} = a^{1/3} \left(\frac{N^{ini}}{\beta + \gamma}\right)^{1/3} \, .$$
Taking into account the probability
distribution (\ref{fa}) of the parameter $a$, we obtain the scaling behavior of
the number of relevant nodes,
\begin{equation} 
p(N_{rel}) = \int_0^\infty da
f(a)\mathcal{C}_a\left(\frac{N_{rel}a^{-1/3}}{(N^{ini}/(\beta+\gamma))^{1/3}}\right)
¸\left(\frac{\beta+\gamma}{aN^{ini}}\right)^{1/3}\, .
\end{equation} 
The error made by taking the upper limit of the integral to infinity
vanishes for $N^{ini} \to \infty$.  We introduce the scaling variable
\begin{equation}
z=\frac{N_{rel}}{\left(\frac{N^{ini}}{\beta+\gamma}\right)^{1/3}}\,
,
\end{equation} 
which has then the following probability distribution
\begin{equation} 
P(z) = \int_0^\infty da \frac {f(a)}{a^{1/3}} \mathcal{C}_a
\left(\frac{z}{a^{1/3}}\right)\, .  
\end{equation}
 The probability distribution for the number of relevant nodes depends
for large $N^{ini}$ only on the scaling variable $z$.  We determined
numerically the function $P(z)$ by combining the two stochastic
processes described in this paper. First, we determined a value of $a$
using the process of Section \ref{fluctuations}. Then, we used this
value of $a$ to determine the last exit time of the stochastic process
of this section, giving a value of $z$. The shape of the curves $P(z)$
depends on the value of $\gamma/\beta$, and the results are shown in
Fig.~\ref{fig5} for $\gamma/\beta=0$ and $\gamma/\beta = 4$, which is
the original Kauffman model, where each update function has the same
weight. It is easy to check analytically that $$\lim_{z\to 0} P(z) =
\sqrt{2\pi}/4(1+\gamma/\beta)^{1/3}\, .$$

The mean number of relevant nodes is
\begin{equation}
\bar{N}_{rel}=\int_0^\infty N_{rel} p(N_{rel}) dN_{rel} =
\left(\frac{N^{ini}}{\beta+\gamma}\right)^{1/3} \int_0^\infty z
P(z) dz\, , 
\end{equation}
i.e., it is proportional to $(N^{ini})^{1/3}$. 
Finally, let us give the probability distribution for the number of
relevant nodes with two relevant inputs. Let $m$ denote the number of
relevant nodes with two relevant inputs and $\tilde P(m;z)dz$ the
probability of having the number of relevant nodes in the interval
$[N_{rel}(z),N_{rel}(z+dz)]$, with $m$ of them having two relevant
inputs. Using Equations (\ref{po}) and (\ref{pi}), we can express
$\tilde P$ as
\begin{eqnarray}
\tilde P(m;z) &=& \int_0^\infty da \frac {f(a)}{a^{1/3}} \mathcal{C}_a
\left(\frac{z}{a^{1/3}}\right)
\nonumber\\
&&\quad \times \frac{P_o\left(m|za^{-1/3}\right)P_i\left(m|za^{-1/3}\right)}{\sum_l P_o\left(l|za^{-1/3}\right)P_i\left(l|za^{-1/3}\right)}\, .\nonumber\\
\end{eqnarray}
As $ P_o$ and $P_i$ decay exponentially fast with increasing $m$, the
mean number of relevant nodes with two inputs is finite. 

\section{Conclusions}

In this paper, we have obtained the asymptotic probability
distributions in the limit of large network size for the number of
nonfrozen nodes, the number of nonfrozen nodes with two nonfrozen
inputs, the number of relevant nodes, and the number of relevant nodes
with two relevant inputs. The mean values of these quantities scale
with network size $N^{ini}$ as a power law in $N^{ini}$, with the
exponent being $2/3$, $1/3$, $1/3$, and $0$ respectively. The
implications of the results are manifold.

First, the notion that these networks are ``critical'' is now
corroborated by the existence of power laws and scaling
functions. Originally, it was expected that the quantities that
display the scaling behavior should be the attractors of the network
\cite{kauffman:metabolic}. In the meantime, it has become clear that
mean attractor numbers do not obey power laws
\cite{samuelsson:superpolynomial}. It is the number of  nonfrozen and
relevant nodes that show scaling behavior.

Next, let us compare the results to those of critical $K=1$ networks. A $K=1$
critical network with $N$ nodes corresponds to the nonfrozen part of a
critical $K=2$ network for $a=0$. In
this case, the probability distribution of the number of relevant
nodes is given by Eq.~(\ref{pa}) with $a=0$, 
\begin{equation}
p_0(N_{rel}) = \frac 1 {\sqrt{N}} \mathcal{C}_0\left(\frac{N_{rel}}{\sqrt{N}}\right)
= \frac{N_{rel}}{N}e^{-N_{rel}^2/2N}\, .
\end{equation}
The mean number of relevant nodes is proportional to $\sqrt{N}$. When
these relevant nodes are connected to a network by pairwise connecting
the inputs and outputs, one obtains a set of simple loops. From
\cite{drossel:number}, we know that there is  a mean number of
$\ln\sqrt{N}$ loops and that the number of loops of length $l$
in a critical $K=1$ network is Poisson distributed with a mean $1/l$
for $l \ll \sqrt{N}$. 
This can be easily explained by consindering the
process of connecting inputs and outputs: We begin with a given node
and draw the node that provides its input from all possible
nodes. Then, we draw the node that provides the input to the newly
chosen node, etc., until the first node is chosen and a loop is
formed. For small loop size, the probability that the loop is closed
after the addition of the $l$th node is $1/N_{rel}$.  Therefore, the
probability that a given node is on a loop of size $l$ is $1/N_{rel}$,
and the mean number of nodes on loops of size $l$ is 1, and the number
of loops of length $l$ is Poisson distributed with a mean $1/l$ for
sufficiently small $l$.

Now, the $K=2$ critical networks have of the order of
$(N^{ini})^{1/3}$ relevant nodes, with only a finite number of them
having two relevant inputs.  The relevant components are constructed
from the relevant nodes by pairwise connecting inputs and outputs.  In
the asymptotic limit of very large $N^{ini}$ that we are considering,
the probability that a randomly chosen relevant node has two inputs or
two outputs vanishes. Let us again construct a component by starting
with one node and choosing its input node etc., until the component is
finished. If the component is small, it consists almost certainly only
of nodes with one input and one output and is therefore a simple
loop. There is no difference between the statistics of the small
relevant components of a $K=1$ critical network, and the number of
loops of length $l$ is Poisson distributed with a mean $1/l$. The
total number of relevant nodes in loops of size $l\le l_c$ with $l_c =
\epsilon (N^{ini})^{1/3}$ (with a small $\epsilon$) is $l_c$, and it
is a small proportion of all nodes. If there were no nodes with two
inputs or outputs, the number of components larger than $l_c$ would be
$(\ln N_{rel} - \ln l_c) = \ln(1/\epsilon)$. The additional links may
reduce this number, which is in any case finite. Since these large
components contain almost all nodes, they contain almost all relevant
nodes with two inputs or outputs.

From these findings, we can obtain results for the attractors of $K=2$
critical networks. The numbers and lengths of attractors are
determined by the relevant components. We now argue that the mean
number and length of attractors increases faster than any power
law. If we remove the components of size larger than $l_c$ and
determine the mean number and length of attractors for this reduced
relevant network, we have a lower bound to the correct numbers. Now,
the reduced relevant network of a $K=2$ system is identical to that of
a critical $K=1$ system (where the critical loop size is $l_c =
\epsilon \sqrt{N}$). In \cite{drossel:number}, it was proven that the
mean number and length of attractors for such a reduced $K=1$ system
increases faster than any power law with network size. We therefore
conclude that the same is true for critical $K=2$ networks. 

Earlier, Samuelsson and Troein \cite{samuelsson:superpolynomial} have
derived analytically an exact expression for the number of attractors
of length $L$ of a critical $K=2$ network in the limit of large
$N^{ini}$, and they have pointed out that this implies that the mean
number of attractors increases faster than any power law with
$N^{ini}$. Using their calculation, it has recently been shown \cite{drossel:onnumber} that
there is a close relationship between $K=1$ critical networks and the
nonfrozen part of $K=2$ critical networks, and that the results of
\cite{samuelsson:superpolynomial} can be most naturally interpreted if
the relevant components of these two networks look identical for
component sizes below the above-given cutoffs. This interpretation is
placed on a firm foundation by the present paper.

\bibliography{fgBib4Tamara.bib}

\begin{thebibliography}{13}
\expandafter\ifx\csname natexlab\endcsname\relax\def\natexlab#1{#1}\fi
\expandafter\ifx\csname bibnamefont\endcsname\relax
  \def\bibnamefont#1{#1}\fi
\expandafter\ifx\csname bibfnamefont\endcsname\relax
  \def\bibfnamefont#1{#1}\fi
\expandafter\ifx\csname citenamefont\endcsname\relax
  \def\citenamefont#1{#1}\fi
\expandafter\ifx\csname url\endcsname\relax
  \def\url#1{\texttt{#1}}\fi
\expandafter\ifx\csname urlprefix\endcsname\relax\def\urlprefix{URL }\fi
\providecommand{\bibinfo}[2]{#2}
\providecommand{\eprint}[2][]{\url{#2}}

\bibitem[{\citenamefont{Kauffman et~al.}(2003)\citenamefont{Kauffman, Peterson,
  Samuelsson, and Troein}}]{kauffman:random}
\bibinfo{author}{\bibfnamefont{S.}~\bibnamefont{Kauffman}},
  \bibinfo{author}{\bibfnamefont{C.}~\bibnamefont{Peterson}},
  \bibinfo{author}{\bibfnamefont{B.}~\bibnamefont{Samuelsson}},
  \bibnamefont{and} \bibinfo{author}{\bibfnamefont{C.}~\bibnamefont{Troein}},
  in \emph{\bibinfo{booktitle}{Proceedings of the National Academy of Sciences
  USA}} (\bibinfo{year}{2003}), no.~\bibinfo{number}{25} in
  \bibinfo{series}{100}, pp. \bibinfo{pages}{14796--14799}.

\bibitem[{\citenamefont{Kauffman}(1969)}]{kauffman:metabolic}
\bibinfo{author}{\bibfnamefont{S.~A.} \bibnamefont{Kauffman}},
  \bibinfo{journal}{J. Theor. Biol.} \textbf{\bibinfo{volume}{22}},
  \bibinfo{pages}{437} (\bibinfo{year}{1969}).

\bibitem[{\citenamefont{Derrida and Pomeau}(1986)}]{derrida:random}
\bibinfo{author}{\bibfnamefont{B.}~\bibnamefont{Derrida}} \bibnamefont{and}
  \bibinfo{author}{\bibfnamefont{Y.}~\bibnamefont{Pomeau}},
  \bibinfo{journal}{Europhys. Lett.} \textbf{\bibinfo{volume}{1}},
  \bibinfo{pages}{45} (\bibinfo{year}{1986}).

\bibitem[{\citenamefont{Derrida and Stauffer}(1986)}]{derrida:phase}
\bibinfo{author}{\bibfnamefont{B.}~\bibnamefont{Derrida}} \bibnamefont{and}
  \bibinfo{author}{\bibfnamefont{D.}~\bibnamefont{Stauffer}},
  \bibinfo{journal}{Europhys. Lett.} \textbf{\bibinfo{volume}{2}},
  \bibinfo{pages}{739} (\bibinfo{year}{1986}).
 

\bibitem[{\citenamefont{Aldana-Gonzalez
  et~al.}(2003)\citenamefont{Aldana-Gonzalez, Coppersmith, and
  Kadanoff}}]{aldana-gonzalez:boolean}
\bibinfo{author}{\bibfnamefont{M.}~\bibnamefont{Aldana-Gonzalez}},
  \bibinfo{author}{\bibfnamefont{S.}~\bibnamefont{Coppersmith}},
  \bibnamefont{and} \bibinfo{author}{\bibfnamefont{L.~P.}
  \bibnamefont{Kadanoff}}, \bibinfo{journal}{Perspectives and Problems in
  Nonlinear Science} pp. \bibinfo{pages}{23--89} (\bibinfo{year}{2003}).
 

\bibitem[{\citenamefont{Flyvbjerg}(1988)}]{flyvbjerg:order}
\bibinfo{author}{\bibfnamefont{H.}~\bibnamefont{Flyvbjerg}},
  \bibinfo{journal}{J. Phys. A} \textbf{\bibinfo{volume}{21}},
  \bibinfo{pages}{L955} (\bibinfo{year}{1988}).
 

\bibitem[{\citenamefont{Flyvbjerg and Kj\ae{}r}(1988)}]{flyvbjerg:exact}
\bibinfo{author}{\bibfnamefont{H.}~\bibnamefont{Flyvbjerg}} \bibnamefont{and}
  \bibinfo{author}{\bibfnamefont{N.~J.} \bibnamefont{Kj\ae{}r}},
  \bibinfo{journal}{J. Phys. A} \textbf{\bibinfo{volume}{21}},
  \bibinfo{pages}{1695} (\bibinfo{year}{1988}).
 

\bibitem[{\citenamefont{Bastolla and
  Parisi}(1998{\natexlab{a}})}]{bastolla:relevant}
\bibinfo{author}{\bibfnamefont{U.}~\bibnamefont{Bastolla}} \bibnamefont{and}
  \bibinfo{author}{\bibfnamefont{G.}~\bibnamefont{Parisi}},
  \bibinfo{journal}{Physica D} \textbf{\bibinfo{volume}{115}},
  \bibinfo{pages}{203} (\bibinfo{year}{1998}{\natexlab{a}}).
 

\bibitem[{\citenamefont{Bastolla and
  Parisi}(1998{\natexlab{b}})}]{bastolla:modular}
\bibinfo{author}{\bibfnamefont{U.}~\bibnamefont{Bastolla}} \bibnamefont{and}
  \bibinfo{author}{\bibfnamefont{G.}~\bibnamefont{Parisi}},
  \bibinfo{journal}{Physica D} \textbf{\bibinfo{volume}{115}},
  \bibinfo{pages}{219} (\bibinfo{year}{1998}{\natexlab{b}}).
 

\bibitem[{\citenamefont{Socolar and Kauffman}(2003)}]{socolar:scaling}
\bibinfo{author}{\bibfnamefont{J.~E.~S.} \bibnamefont{Socolar}}
  \bibnamefont{and} \bibinfo{author}{\bibfnamefont{S.~A.}
  \bibnamefont{Kauffman}}, \bibinfo{journal}{Phys. Rev. Lett.}
  \textbf{\bibinfo{volume}{90}}, \bibinfo{pages}{068702}
  (\bibinfo{year}{2003}). 

\bibitem[{\citenamefont{Samuelsson and
  Troein}(2003)}]{samuelsson:superpolynomial}
\bibinfo{author}{\bibfnamefont{B.}~\bibnamefont{Samuelsson}} \bibnamefont{and}
  \bibinfo{author}{\bibfnamefont{C.}~\bibnamefont{Troein}},
  \bibinfo{journal}{Phys. Rev. Lett.} \textbf{\bibinfo{volume}{90}},
  \bibinfo{pages}{098701} (\bibinfo{year}{2003}).

\bibitem[{\citenamefont{Drossel}(2005)}]{drossel:onnumber}
\bibinfo{author}{\bibfnamefont{B.}~\bibnamefont{Drossel}},
  \bibinfo{journal}{Phys. Rev. E} \textbf{\bibinfo{volume}{72}},
  \bibinfo{pages}{xxx} (\bibinfo{year}{2005}).

\bibitem[{\citenamefont{Drossel et~al.}(2005)\citenamefont{Drossel, Mihaljev,
  and Greil}}]{drossel:number}
\bibinfo{author}{\bibfnamefont{B.}~\bibnamefont{Drossel}},
  \bibinfo{author}{\bibfnamefont{T.}~\bibnamefont{Mihaljev}}, \bibnamefont{and}
  \bibinfo{author}{\bibfnamefont{F.}~\bibnamefont{Greil}},
  \bibinfo{journal}{Phys. Rev. Lett.} \textbf{\bibinfo{volume}{94}},
  \bibinfo{pages}{088701} (\bibinfo{year}{2005}).

\end{thebibliography}

\end{document}